# Symbolic Execution for Deep Neural Networks

Divya Gopinath[1], Kaiyuan Wang[2], Mengshi Zhang[2], Corina S. Păsăreanu[1], Sarfraz Khurshid[2]
[1] Carnegie Mellon University Silicon Valley, Moffett Field, CA 94035, USA
{divyag1@andrew.,pcorina}@cmu.edu
[2] University of Texas at Austin, TX 78712, USA
{kaiyuanw,mengshi.zhang,khurshid}@utexas.edu

*Abstract*—Deep Neural Networks (DNN) are increasingly used in a variety of applications, many of them with substantial safety and security concerns. This paper introduces DeepCheck, a new approach for validating DNNs based on core ideas from program analysis, specifically from *symbolic execution*. The idea is to translate a DNN into an imperative program, thereby enabling program analysis to assist with DNN validation. A basic translation however creates programs that are very complex to analyze. DeepCheck introduces novel techniques for lightweight symbolic analysis of DNNs and applies them in the context of image classification to address two challenging problems in DNN analysis: 1) identification of *important* pixels (for attribution and adversarial generation); and 2) creation of 1-pixel and 2-pixel attacks. Experimental results using the MNIST data-set show that DeepCheck's lightweight symbolic analysis provides a valuable tool for DNN validation.

## I. INTRODUCTION

Deep Neural Networks (DNN) are increasingly used in a variety of applications, many of them with substantial safety and security concerns [19]. Our focus in this paper is on *image classifiers*: DNNs that take in complex, high dimensional input, pass it through multiple layers of transformations, and finally assign to it a specific output label. Such networks are now being integrated into the perception modules of autonomous or semi-autonomous vehicles, at major car companies such as Tesla, BMW, Ford, and others. It is expected that this trend will continue and intensify, with neural networks being increasingly used in safety critical applications which require high assurance guarantees.

Thus, the traditional emphasis on obtaining high accuracy for DNNs is being augmented with safety and security goals [15]. However, validating DNNs is complex and challenging, due to the nature of the learning techniques that create these models. For example, it is not well understood *why* a DNN, say an image classifier, gives a particular output. This inability to explain the DNN decisions hinders their application in safety critical domains, such as autonomy. Furthermore, evaluating the robustness of a network, e.g., against conceptually simple yet effective attacks, such as 1-pixel attacks, where just one pixel on an image is altered to make the network classify the image incorrectly, is a hard technical problem, due to the huge input space of all images.

This paper presents an approach for the analysis of deep neural networks based on *symbolic execution* [7], [18]. Symbolic execution is a well-known program analysis technique that has seen many advances in recent years [3], [4], [13], [17], [24] and applications in various domains, such as security [6], [8], smartphone apps [1], operating systems [31], and databases [11].

Traditional symbolic execution executes programs on *symbolic*, instead of concrete, inputs and systematically explores the program paths (up to a given depth bound). For each path explored, it builds *path conditions*, i.e., constraints on program inputs that execute that path based on the conditional branches in the code. To illustrate, when a conditional statement, say "$if(c)...$" is executed, each of the two conditional branches is individually explored, and the path condition $PC$ is updated to $PC \wedge c$ for the *then* branch and to $PC \wedge \neg c$ for the *else* branch. The feasibility of the path conditions is checked using off-the-shelf constraint solvers, such as satisfiability modulo theories (SMT) solvers [2], [9], as branch conditions are encountered during symbolic execution to detect and avoid infeasible paths (if possible) and to generate test inputs that execute feasible paths (as desired). Overall, the program effects are computed as *functions* over the symbolic inputs.

Symbolic execution for deep neural networks is attractive because it would allow us to extract mathematical characterizations (in the form of path conditions and symbolic expressions) of the *internal* behavior of the networks, which are notoriously opaque. The information computed with symbolic execution – even without constraint solving – can be used to examine the *coverage* of the neural network, to extract *explanations* of behavior, and to identify pixels *sensitive to adversarial* behavior. Symbolic execution with constraint solving can also be used to generate *new inputs* to test the networks and find new vulnerabilities.

However, there are several challenges to developing effective symbolic execution for neural networks: (1) typically the networks have no branching; (2) the networks in general are highly non-linear, and solvers for constraints that ensue for such systems are not well-developed; and (3) there are serious scalability issues: state-of-the art neural networks consist of thousands of neurons that are well beyond the capabilities of current symbolic reasoning tools.

Our idea is that a class of neural networks can be translated to imperative programs that can feasibly be analyzed using core ideas from symbolic execution. Specifically, we apply symbolic execution to networks that use the rectified linear units (ReLUs) activation functions. These activation functions naturally admit a branching structure of the form "$if$ $(x > 0)$ $... else ...$", and thus, a path through the neural network can conceptually be viewed as a path through the translated program. Hence, path conditions for paths through the network can be built using symbolic execution of the corresponding program, and analyzed as needed.

We apply symbolic execution to select program paths of interest, e.g., paths taken by specific input images, say from the network's training data. However, just building the path condition for even one path, say using a straightforward application of concolic (or dynamic symbolic) execution [3], [4], [13], [24], can take considerable amount of time, and solving a path condition with just one symbolic variable can stress modern SMT solvers due to the parsing and simplifications the solver must do to handle constraints, which are conceptually rather simple but syntactically quite complex. We introduce efficient techniques for building path conditions and utilizing solvers. Our approach embodies a lightweight yet particularly insightful analysis that offers insights into the *reasoning* performed by the networks to make classification decisions with no constraint solving at all, and allows a directed approach for creating adversarial images for network validation with minimal constraint solving.

This paper makes the following contributions:

- **Idea.** We introduce the idea of using symbolic execution for analyzing neural networks by translating them to imperative code that is practical to analyze, specifically focusing on important pixel identification, and 1-pixel and 2-pixel attack generation.
- **Approach.** We introduce the DeepCheck approach that is based on translating neural networks with rectified linear units to analyzable imperative code and embodied by two validation techniques; DeepCheck$^{Imp}$, which applies symbolic execution for identifying important pixels that intuitively provide explanations for why neural networks make certain decision; and DeepCheck$^{tPA}$, which applies symbolic execution to create 1-pixel and 2-pixel attacks.
- **Evaluation.** We present an experimental evaluation to address three key research questions using the MNIST dataset that consists of images of numeric digits and has been widely studied in literature. The experimental results show that it is feasible to use symbolic execution to identify important pixels and to create 1-pixel and 2-pixel attacks, and that important pixels enable a more scalable approach for generating 1-pixel and 2-pixel attacks. For example, for images of 9 out of 10 digits, a 2-pixel attack is found by checking the 2-pixel combinations of just top-4 important pixels, i.e., a pair of pixels to attack the network is found by checking no more than 6 pixel pairs.

## II. BACKGROUND: NEURAL NETWORKS

Neural networks are often used as *classifiers*, meaning that they assign to each input an output label/class. Such a neural network $F$ can be regarded as a function that assigns to input $x$ an output label $y$, denoted as $F(x) = y$. Internally, a neural network is comprised of multiple layers of nodes, called neurons. Each node refines and extracts information from values computed by nodes in the previous layer. The first layer is the *input* layer, which takes in the input variables (also called features). The network may have several *hidden* layers: each of its neurons computes a weighted sum of the input variables according to a unique weight vector and a bias value, and then applies an *activation function* to the result $(h_{W,b}(x) = f(\sum_{i=1}^{3} W_i x_i + b))$. Most recent networks use rectified linear units (ReLUs) activation functions. A rectified linear unit has output 0 if the input is less than 0, and raw output otherwise; $f(x) = max(x, 0)$. The last layer uses a *softmax* function to assign an the output class is the input. The softmax function squashes the outputs of each node of the previous layer to be between 0 and 1, equivalent to a categorical probability distribution. The number of nodes in this layer is equal to the number of output classes and their respective outputs gives the probability of the input being classified to that class.

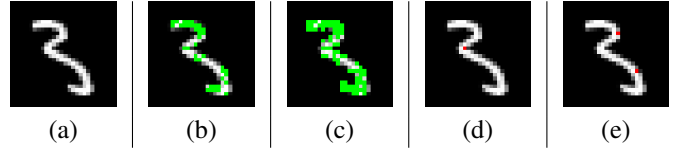

Fig. 1: (a) Example image with predicted label 3. (b) Top-5% important pixels (highlighted in green) identified by DeepCheck$^{Imp}$. (c) Top-10% important pixels (green) identified by DeepCheck$^{Imp}$. (d) 1-pixel attack (highlighted in red) identified by DeepCheck$^{tPA}$; changing the red-pixel to black changes the predicted label to 8. (e) 2-pixel attack (red) that does include an attackable pixel for 1-pixel attack.

## III. OVERVIEW

This section gives an illustrative overview of our approach to demonstrate our reduction of deep neural nets to imperative code and use of symbolic execution. Our subject neural network $\mathcal{N}$ is a fully connected $784 \times 10 \times 10 \times 10$ network, which has been trained on all 60,000 images in the training data of the MNIST dataset [20], and has an accuracy of 92%.

Given the trained network $\mathcal{N}$, we apply our technique DeepCheck$^\tau$ to translate it to an imperative program $\mathcal{P}$ that has the same behavior as the original network but is amenable to program analysis. Figure 1(a) shows an example image $\mathcal{I}$ from the standard MNIST training data, which has the predicted label of 3 (which is the same as its true label).

### A. Identifying important pixels

Given $\mathcal{I}$ as an input, our important pixel identification technique DeepCheck$^{Imp}$ executes the program $\mathcal{P}$, and for the execution path taken by $\mathcal{I}$, computes for every output label, a linear expression in terms of the input variables which are the 784 pixels of the input image. The algorithm then uses the coefficients of the input pixels in the expression corresponding to the label assigned by the network (3 in the case of the example), to assign an *importance score* for every pixel. A pixel $p_1$ is considered more important than another $p_2$, if the classification decision is impacted more by $p_1$ than $p_2$. DeepCheck$^{Imp}$ employs three metrics; *abs*, *co*, *coi*, to calculate the importance of each pixel. The pixels are then sorted in the descending order of their scores. The pixels which are higher on this list (top threshold %) are identified as being *important*. The insight is that the short-listed important pixels can be held responsible for the classification decision or the classification can be explained or attributed to them. Therefore, a small change to the image with respect to the important pixels, such as changing the value of just one important pixel can have a high impact on the classification decision, and sometimes may lead to the discovery of *adversarial* examples – the new image differs from the original image by the value of just one pixel but this makes the network incorrectly assign a different label to this image.

Figure 1(b) illustrates the top-5%, i.e., 39, important pixels highlighted in green. Note, how the important pixels trace the shape of the digit 3 and do not point to areas of the image irrelevant to the digit being identified as 3 such as the background or the edges. Figure 1(c) illustrates the top-10%, i.e., 78, important pixels highlighted in green. These important pixels form a denser pattern that traces the shape of the digit 3. This highlights that short-listing pixels based on their importance score in terms of the coefficients of the expression for the expected label, enables the identification of pixels that can *explain* the classification decision.

### B. Identifying attack pixels

Conceptually, our $t$-pixel attack technique DeepCheck$^{tPA}$ aims to create a new image that differs from the original image at $t$ pixels, and has (1) the same *activation pattern* as the original image but (2) a different label from the original image. Specifically, for a 1-pixel attack, DeepCheck$^{tPA}$ selects a pixel $p$, makes its value symbolic $p_s$, retains the original concrete values for all other pixels, and constructs a constraint-solving problem, which requires (1) execution of the same path up to the output layer as the original image $\mathcal{I}$ and (2) change in the output label from the original predicted label of $\mathcal{I}$. The constraint-solving problem consists of a simplified path condition for image $\mathcal{I}$'s execution path such that the path condition contains only one symbolic value, i.e., $p_s$. If the constraint is satisfiable, a solution provides the value for pixel $p$ to create the 1-pixel attack image. For solving constraints, we use the SMT solver Z3 [9].

DeepCheck$^{tPA}$ checks whether any pixel of $\mathcal{I}$ can be attacked by making one pixel symbolic at a time and checking the resulting path condition. Figure 1(d) shows a 1-pixel attack identified by our approach for image $\mathcal{I}$; changing the red pixel to black changes the predicted label of the image to 8. This attackable pixel actually lies in the top-5% (top 39) important pixels for $\mathcal{I}$ identified by DeepCheck$^{Imp}$. The rank order of this attackable pixel in descending order of importance is 21. For this case, focusing the 1-pixel attack on important pixels can allow finding an attack much quicker than checking every pixel for attackability. In fact, this image only has one 1-pixel attack. A linear search that starts at the first image pixel (top-left corner) and scans left-to-right takes 346 attempts to find this attack pixel, which is over 16X the attackable pixel's rank-order (21). We believe *important pixels* can provide a practical heuristic for a *more scalable approach to create attacks*.

To create 2-pixel attacks, we focus DeepCheck$^{tPA}$ on the important pixels identified by DeepCheck$^{Imp}$, specifically on the top-5% important pixels. We make $\binom{39}{2}$ = 741 unordered pairs of the selected important pixels, and for each pair, we make the two corresponding variables symbolic, so each path condition created by symbolic execution contains exactly two symbolic variables. Applying DeepCheck$^{tPA}$ to the 741 pairs results in 93 unique 2-pixel attacks. 38 of the 2-pixel attack pairs contain as an element the pixel that was earlier identified for the 1-pixel attack, whereas 55 of the pairs contain only pixels that are not 1-pixel attackable; Figure 1(e) shows one such pair in red.

For this example the important pixels identified by DeepCheck$^{Imp}$ play a key role in focusing DeepCheck$^{tPA}$ to find a 2-pixel attack. The first attack found by DeepCheck$^{tPA}$ includes the 2 of the 3 top-most important pixels. Thus, the search for a 2-pixel attack for this example requires checking no more than just $\binom{3}{2}$ = 3 pairs.

These results illustrate the potential of using symbolic execution in identifying important pixels and creating 1-pixel and 2-pixel attacks, as well as the value of important pixels in finding attackable pixels and pixel-pairs.

## IV. TECHNIQUE

We build a symbolic execution framework that i) translates a given neural network into a semantically equivalent imperative program, ii) performs symbolic execution to collect constraints on paths for specific concrete inputs, iii) identifies input features or pixels important to the classification, and iv) synthesizes adversaries that fool the network into misclassification.

### A. Symbolic Execution Framework

The input to our framework is a feed-forward neural network with linear activation functions. Consider network $NN$ with $m$ inputs $X :< x_0, ..., x_{m-1} >$, $n_l$ layers with $l$ representing each layer, the weights being represented as $w_{i,j}^l$, i.e., the weight of the edge connecting the node $j$ in layer $l$ to node $i$ in layer $l+1$, $b_j^l$ representing the bias term added to the weighted sum of the input variables from layer $l$ to the node $j$ in layer $l+1$. Let $Y :< y_0, ..., y_{n-1} >$ be the output of the nodes before the application of softmax where $n$ is the number of output labels. We consider the second last layer to be the output layer of our model. The output of the classifier is calculated as the index containing the maximum value at the output layer. The goal is to translate $NN$ to an imperative program $\mathcal{P}$ such that for any input $x$, $NN(x) = \mathcal{P}(x)$. The input state of $\mathcal{P}$ is $S_{inp} =< s_{i_0}, ..., s_{i_{m-1}} >$, where each $s_i$ corresponds to an input variable $x_i$, the output state is $S_{op} =< s_{o_0}, ..., s_{o_{n-1}} >$, where each $s_o$ corresponds to $y_i$, and the hidden nodes represent the intermediate states $S^h =< s_0^h, ..., s_{x-1}^h >$ where $h$ is the hidden layer number and $v^h$ is the number of neurons in layer $h$.

*a) Translation:* A typical neural network structure does not have any branching. However, observe that in the case of rectified linear units, the activation function $f(x) = max(0, x)$ can be naturally translated into a branching instruction, *if* $(x > 0)$ *then return* $x$; *else return* 0;. Thus, a path through the neural network can be seen as a path through the translated program, where each executed branch corresponds to the neural network node being activated or not. Every neuron, which is the basic computational unit of the network, applies the function, $h_{W,b}(x)$ on its input $x$ (section II, where $x$ is

```
for (index = 0; index < v^{h-1}; index++) {
    val += w^{h-1}_{index,i} * s^{h-1}_{index} ; }
val += b_i;
if (val > 0) { s^h_i = val; }
else { s^h_i = 0; }
```

Fig. 2: Code-snippet for Neuron $i$.

```
for (index = 0; index < v^{h-1}; index++) {
    val += w^{h-1}_{index,i} * s^{h-1}_{index} ;
    coef_{temp} += ( w^{h-1}_{index,i} · coef^{h-1}_{index} ); }
val += b_i;
if (val > 0) { s^h_i = val; coef^h_i = coef_{temp}; }
else { s^h_i = 0; coef^h_i = coef_{temp} · 0; }
```

Fig. 3: operations for $coef$ array represent element vise addition and multiplication ($\cdot$ represents dot-product).

the outputs of the nodes in the previous layer respectively. The code-snippet corresponding to a node $i$ in layer $h$ is shown in Figure 2, where $v^{h-1}$ represents the number of nodes at layer $h-1$. This code is invoked $v$ times for every node of the layer $h$. The corresponding outputs are fed as inputs to the subsequent layer, until the layer before softmax to calculate $S_{op} = <s_{o_0}, ..., s_{o_{n-1}}>$.

*b) Analysis:* Once we translate a given $NN$ to a corresponding semantically equivalent $\mathcal{P}$, we perform execution in concolic mode on a given concrete input. The sections below describe this process and explain how it is used to identify important input attributes and synthesize attacks using constraint solving.

### B. Identifying important pixels

This section describes in the detail the algorithm of DeepCheck$^{Imp}$ to short-list *important pixels* that could act as *explanations* for the classification decision and also identify pixels *vulnerable to adversaries*.

The function representing a multiple-layer neural network could be expressed in terms of the weights and biases for every layer. Consider the network $NN$ as defined earlier. Let $a^h_j$ represent the intermediate outputs of the hidden nodes in layer $h$ and node $j$. For instance, for a three layer network, presented below would be the representation of the output in terms of the weights and biases of the network.

$a^1_0 = f(w^0_{0,0} * x_0 + w^0_{0,1} * x_1 + w^0_{0,2} * x_2 + b^0_0)$
$a^1_1 = f(w^0_{1,0} * x_0 + w^0_{1,1} * x_1 + w^0_{1,2} * x_2 + b^0_1)$
$a^1_2 = f(w^0_{2,0} * x_0 + w^0_{2,1} * x_1 + w^0_{2,2} * x_2 + b^0_1)$
$a^2_0 = y_0 = f(w^1_{0,0} * a^1_0 + w^1_{0,1} * a^1_1 + w^1_{0,2} * a^1_2 + b^1_0)$
$a^2_1 = y_1 = f(w^1_{1,0} * a^1_0 + w^1_{1,1} * a^1_1 + w^1_{1,2} * a^1_2 + b^1_1)$

In the case of a network having linear activations such as $f(x)$ being a ReLU function, the network is a linear model. Each element of the $Y$ could be expressed in the form of a linear polynomial expression in terms of the input variables. For instance, in the previous equation for $y_0$, we could remove the function term and replace the values of $a^h_j$ with their respective equations in terms of the input variables (assuming each $a^h_j$ evaluates to a value greater than or equal to 0). $y_0 = w^1_{0,0}*(w^0_{0,0}*x_0+w^0_{0,1}*x_1+w^0_{0,2}*x_2+b^0_0)+w^1_{0,1}*(w^0_{1,0}*x_0+w^0_{1,1}*x_1+w^0_{1,2}*x_2+b^0_1)+w^1_{0,2}*(w^0_{2,0}*x_0+w^0_{2,1}*x_1+w^0_{2,2}*x_2+b^0_1)+b^1_0$. In general in a feed-forward network with ReLU activation functions, each output element, $y_i$, could be expressed as $y_i = C_{i,0} * x_0 + C_{i,1} * x_1 + ... + C_{i,n-1} * x_{n-1}$, where $C_{i,0}$ to $C_{i,n-1}$ are coefficients (signed) of the linear polynomial, that can calculated in terms of the weights of the non-zero edges from $x_j$ to $y_i$.

$$C_{i,j} = \sum_{p=0}^{\text{\# of paths from } x_j \text{ to } y_i} ( \prod_{e=0}^{\text{\# of edges in } p} w(e)) \quad (1)$$

*a) DeepCheck$^{Imp}$ Algorithm::* Given a concrete input value $\mathcal{I}$ and a translation of the neural network into a program $\mathcal{P}$, DeepCheck$^{Imp}$ instruments the code such that execution of the program on the input also simultaneously updates its impact on the input variables, thereby finally computing the coefficients for the input pixels in the output expression corresponding to the classification decision, $y_{label}$. It then uses them to identify the important pixels. Let us consider the program $\mathcal{P}$ as defined before. Each of these state variables are data-dependent on the input variables. Therefore, we maintain a coefficients array of the size of the input variables corresponding to each of the state-variables, $coef^h_{i,j}$, where $h$ is the layer number (0 to the second-last layer), $i$ is the neuron or node number at that layer (0 to $v^h - 1$) and $j$ is the index of the input variable (0 to $m - 1$).

Given a concrete input $\mathcal{I}$, the co-efficient array is initialized as follows $coef^0_{i,j} = 1$ if $i = j$, 0 otherwise. The execution follows the path taken by the concrete input, i.e. path conditions are evaluated as per the input values. Every time an operation is encountered during the execution of the path, the coefficients arrays are updated accordingly. Consider the code-snippet (Fig. 3) showing the operations performed to calculate the concrete value of $s^h_i$ and the corresponding update to coefficients $coef^h_{i,0}, ..., coef^h_{i,m-1}$. Note that since the calculation is done dynamically as the network is being executed on the given input, the co-efficient arrays are updated based on the actual values of the activation functions. For instance, wherever the ReLU activation function evaluates to a zero, the respective co-efficient arrays get reset to zeros.

After the execution of the first hidden layer 1, for each neuron $i$ the coefficients corresponding to each input variable would be $w^0_{0,i}, w^0_{1,i}, ...w^0_{m-1,i}$ respectively. Similarly after the execution of the second hidden layer 2, for each neuron $i$, the coefficients corresponding to each input variable would be $w^1_{0,i}*w^0_{0,0}+w^1_{1,i}*w^0_{0,1}+...+w^1_{v^1,i}*w^0_{0,v^1}, ..., w^1_{0,i}*w^0_{m-1,0}+w^1_{1,i}*w^0_{m-1,1}+...+w^1_{v^1,i}*w^0_{m-1,v^1}$, where the values of $w^0_{0,x}, ..., w^0_{m-1,x}$ would be equivalent to zeros respectively if $s^1_x$ evaluates to zero. Therefore, ultimately at the layer before softmax, the coefficients computed for $y_{label}$ corresponding to the each input variable would be the summation of the products of weights along the non-zero edges from each input variable to $y_{label}$, equivalent to Eq. 1. Note that these coefficients may be different from the one extracted from the entire model with no non-zero edges. The coefficients extracted by DeepCheck$^{Imp}$ precisely corresponds to the non-zero edges of the network for the given input.

**Importance metrics:** Many existing techniques use a gradient-based approach VI to determine the impact of each input variable on specific output variables. In a linear model, the partial derivative of an output variable w.r.t an input variable $dy_i/dx_j$ precisely corresponds to the corresponding co-efficient $C_{i,j}$. Therefore, we use the value of the co-efficient of the input variable to determine its impact on the output variable, akin to gradient based approaches that use the derivative to determine the impact of each input variable. More specifically, we consider the coefficients of the output variable corresponding to the label assigned by the network to the input, $y_{label}$. We calculate an important score for each input variable w.r.t. three metrics presented below;

- *co*: We consider the actual values of the coefficients (signed partial derivative $dy_{label}/dx_j$) to determine the impact of the input variables on $y_{label}$. This metric would assign higher scores to input variables that would impact $y_{label}$ for all inputs semantically equivalent to the given input; trigger the same activations for the network but may have different values for the input variables. For the example Figure 1, the following were the top ten input pixels with the corresponding *co* values $<pixel, value>$; $<742, 1.137>, <489, 1.013>, <488, 1.006>, <434, 0.914>, <286, 0.905>, <458, 0.887>, <666, 0.879>, <408, 0.832>, <377, 0.825>, <383, 0.816>, <739, 0.805>$
- *coi*: In order to identify pixels that can be attributed for the classification decision for the specific given input, we need to consider the input values as well. This can be determined by multiplying the co-efficient values with the corresponding input values, similar to techniques such as DeepLIFT [25] ($x_j * dy_i/dx_j$). These would serve as *better explanations* for the specific input to be assigned the specific label. For the example Figure 1, the following were the top ten input pixels with the corresponding *coi* values $<pixel, value>$; $<434, 1.365>, <405, 1.076>, <318, 0.901>, <292, 0.890>, <662, 0.881>, <237, 0.826>, <433, 0.788>, <209, 0.776>, <403, 0.773>, <463, 0.695>, <177, 0.625>$
- *abs*: The absolute values of the coefficients for the expression corresponding to $y_{label}$ represent the magnitude of the impact that a change in the respective input variable would have on the value of $y_{label}$, irrespective of whether it causes an increase or decrease. Therefore, akin to saliency maps, absolute value of coefficients could aid in identifying input variables that the classification is most vulnerable to. A small change in the value of these variables can lead to the decision changing thus potentially identifying *adversarial examples*. For the example Figure 1, the following were the top ten input pixels with the corresponding *abs* values $<pixel, value>$; $<742, 1.136>, <489, 1.013>, <488, 1.006>$, $<415, 0.986>, <441, 0.978>, <434, 0.914>, <286, 0.905>, <482, 0.899>, <458, 0.887>, <666, 0.879>, <293, 0.872>$

The input variables are ordered in descending order of their importance scores and the pixels corresponding to the top threshold % of the scores are short-listed as being *important*. The value of the threshold can be user-defined. We have experimented with 5, 10 and 30.

### C. Synthesizing Attacks

For a given input image $\mathcal{I}$, DeepCheck$^{tPA}$ sets $t$ pixels with symbolic values and the rest pixels with concrete values. Then, it collects path condition of $\mathcal{I}$ for the neural network program $\mathcal{P}$ right before the softmax layer, i.e. layer $Y$. The final path condition is a conjunction of inequalities (introduced by the ReLU function) of the form $PC = \bigwedge_{h=1}^{H} (B^h + \sum_{i=1}^{t} C_i^h \cdot X_i \ \gamma \ 0)$, where $h$ represents the $h^{th}$ activation function defined by the computation order, $H$ is the total number of activation functions. In our fully connected network $\mathcal{N}$, each hidden neuron corresponds to an ReLU activation function and we use the same computation order of the activation function to refer hidden neurons. $B^h$ is the bias term of the output value of the $h^{th}$ hidden neuron; $C_i^h$ and $X_i$ are the $i^{th}$ coefficient and symbolic value of the $h^{th}$ hidden neuron, respectively. $\gamma \in \{>, \leq\}$ and is determined by the activeness of the $h^{th}$ hidden neuron. The path condition determines the neuron activation pattern from the network point of view. In practice, the number of conjunct clauses is smaller than $H$ because sometimes all coefficients of the symbolic values ($C_i^h$) are 0 in which case the entire conjunct clause evaluates to true.

The output value of the $j^{th}$ ($j \in [1, n]$) neuron in layer $Y$ is a function of symbolic values of the form $f_j(X) = B^j + \sum_{i=1}^{t} C_i^j \cdot X_i$. Assume that the network predict label $l$ ($l \in [1, n]$) for the input $\mathcal{I}$, then DeepCheck$^{tPA}$ add constraints $AC = \bigwedge_{j=1, j \neq l'}^{n} f_j(X) < f_{l'}(X)$ to require the network to predict a label $l'$ where $l \neq l'$. Additionally, DeepCheck$^{tPA}$ add constraints $RA = \bigwedge_{i=1}^{t} lo \leq X_i \leq hi$, where $lo$ and $hi$ are the lower and upper bound of input values after data normalization. This is to make sure the solution of $X_i$ is in the range. DeepCheck$^{tPA}$ invokes Z3 with constraints $PC \wedge AC \wedge RC$ to solve for concrete values for all $X_i$. If a solution is found, DeepCheck$^{tPA}$ succeeds in a $t$ pixel attack by setting $X$ with the concrete values Z3 returns and the network predicts label $l'$ which is different from the original predicted label $l$ with the same neuron activation pattern.

## V. EVALUATION

This section describes an experimental evaluation of DeepCheck, specifically the important pixel identification technique DeepCheck$^{Imp}$ and the $t$-pixel attack technique DeepCheck$^{tPA}$. We address three key research questions:

- **RQ1.** Does symbolic execution enable important pixel identification? (Section V-B)

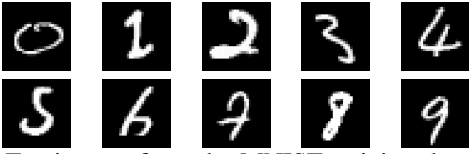

Fig. 4: Ten images from the MNIST training dataset [20].

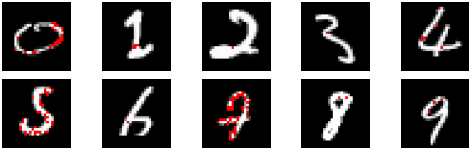

Fig. 5: Attackable pixels for 1-pixel attack highlighted in red.

- **RQ2.** Does symbolic execution enable 1-pixel attacks and 2-pixel attacks? (Section V-C)
- **RQ3.** How do important pixels identified by DeepCheck$^{Imp}$ compare with pixels that can be attacked using DeepCheck$^{tPA}$? (Section V-D)

### A. Subject Neural Network and Images

As our subject neural net we use the image classification network that we described in Section III. Recall, it is a fully connected $784 \times 10 \times 10 \times 10 \times 10$ network, which has been trained on all 60,000 images in the training data of the MNIST dataset [20], and has an accuracy of 92%. We use 10 images from the training data as subject base images for our evaluation; the images cover all 10 labels $0, 1, \ldots, 9$. Figure 4 graphically displays them.

### B. Important pixel identification

For each image, we apply DeepCheck$^{Imp}$ to compute a ranked list of pixels according to their relative importance based on three ranking metrics: *abs*, *co*, *coi*. Table I graphically shows the results produced by DeepCheck$^{Imp}$ for the three metrics for top-5% and top-10% of important pixels. For each image (digit), the table displays the important pixels in green. (Appendix A Table VII displays the results for top-30% of the important pixels with respect to the three metrics.)

For top-5% and top-10% results, each metric generally identifies pixels in the central part of the image as most important, with top-10% forming a denser pattern than top-5%. The *abs* and *co* metrics show similar patterns in the central region. The *coi* metric most closely follows the digit's shape.

Overall, the use of symbolic execution enables identification of important pixels that help explain the classification decisions of the neural network.

### C. t-pixel attack

We apply DeepCheck$^{tPA}$ for 1-pixel attack and for 2-pixel attack. For 1-pixel attack, for each image, we evaluate each of the 784 pixels in the image to determine if it is *attackable*, i.e., can be given a different value to change the image's predicted label while preserving the neuron activation pattern. Figure 5 highlights, for each digit, each attackable pixel (in red) for a 1-pixel attack. The attackable pixels lie on or very close to the shape of the corresponding digit.

TABLE I: Top-5% and top-10% of important pixels (green) identified by DeepCheck$^{Imp}$ for *abs*, *co*, and *coi*.

| digit | 0 | 1 | 2 | 3 | 4 | 5 | 6 | 7 | 8 | 9 |
|---|---|---|---|---|---|---|---|---|---|---|
| #ap | 25 | 4 | 1 | 1 | 6 | 36 | 1 | 47 | 2 | 3 |
| alabel | 5 | 3 | 3 | 8 | 6 | 3 | 5,8 | 9 | 1 | 4 |
| 1stap | 244 | 489 | 516 | 346 | 71 | 103 | 486 | 156 | 211 | 240 |

TABLE II: Number of 1-pixel attackable pixels (#ap) identified by DeepCheck$^{tPA}$ for each image (digit) among all pixels. "alabel" shows the set of labels that the network incorrectly produces after the 1-pixel attacks. "1stap" shows the smallest position (pixel ID) of any attackable pixel in the image. For digit 6, the pixel at position 486 can be attacked to become 5 as well as 8 by appropriately setting the pixel's value.

Table II summarizes the attacks identified by DeepCheck$^{tPA}$ for 1-pixel attack. Some images, e.g., digit 2, contain one attackable pixel out of 784 pixels, whereas some others contain multiple, e.g., 47 for digit 7. All images except digit 6 when attacked get a unique incorrect label (alabel). Digit 6 has 2 attacks but both use the same pixel, which can be attacked in two ways: to incorrect label 5 and incorrect label 8. The row "1stap" shows the smallest position of any attackable pixel for each digit; this position represents the number of attempts DeepCheck$^{tPA}$ takes to find the first attackable pixel for each image when it exhaustively checks every pixel in the image. For some digits, e.g., 4, less than 10% of the pixels are checked. For some other digits, e.g., 1, over 62% of the pixels are checked. Note, this exhaustive search is without utilizing the important pixel identification, which we utilize next.

For 2-pixel attack, for each image, we select the top-5%

| digit | 0 | 1 | 2 | 3 | 4 | 5 | 6 | 7 | 8 | 9 |
|---|---|---|---|---|---|---|---|---|---|---|
| #a2p | 548 | 198 | 48 | 93 | 260 | 463 | 287 | 651 | 111 | 171 |
| #a2p-new | 60 | 87 | 10 | 55 | 114 | 100 | 186 | 75 | 36 | 96 |
| alabel | 5 | 2,3 | 3 | 8 | 6 | 3 | 1,3,5,8 | 8,9 | 1,2,3 | 4 |

TABLE III: Number of 2-pixel attacks (#a2p) identified by DeepCheck$^{tPA}$ for each image (digit), and of these attacks the number that do not use any pixel that is attackable for a 1-pixel attack (#a2p-new). "alabel" shows the set of labels that each digit can be attacked to incorrectly produce using the neural net. The search for 2-pixel attacks is focused on top-5% important pixels identified by the *coi* metric of DeepCheck$^{Imp}$.

| digit | #ap | 5%(39) | | | 10%(78) | | | 30%(235) | | |
|---|---|---|---|---|---|---|---|---|---|---|
| | | abs | co | coi | abs | co | coi | abs | co | coi |
| 0 | 25 | 12.0 | 28.0 | 64.0 | 28.0 | 44.0 | 68.0 | 80.0 | 64.0 | 68.0 |
| 1 | 4 | 0.0 | 25.0 | 75.0 | 25.0 | 50.0 | 100.0 | 75.0 | 100.0 | 100.0 |
| 2 | 1 | 0.0 | 0.0 | 100.0 | 0.0 | 100.0 | 100.0 | 100.0 | 100.0 | 100.0 |
| 3 | 1 | 0.0 | 0.0 | 100.0 | 0.0 | 0.0 | 100.0 | 0.0 | 100.0 | 100.0 |
| 4 | 6 | 66.7 | 50.0 | 66.7 | 66.7 | 50.0 | 66.7 | 100.0 | 66.7 | 66.7 |
| 5 | 36 | 2.8 | 5.6 | 30.6 | 11.1 | 8.3 | 50.0 | 38.9 | 33.3 | 52.8 |
| 6 | 1 | 100.0 | 100.0 | 100.0 | 100.0 | 100.0 | 100.0 | 100.0 | 100.0 | 100.0 |
| 7 | 47 | 19.1 | 17.0 | 40.4 | 29.8 | 25.5 | 48.9 | 48.9 | 40.4 | 48.9 |
| 8 | 2 | 0.0 | 0.0 | 100.0 | 0.0 | 0.0 | 100.0 | 100.0 | 100.0 | 100.0 |
| 9 | 3 | 0.0 | 0.0 | 66.7 | 0.0 | 0.0 | 66.7 | 66.7 | 66.7 | 66.7 |

TABLE IV: Important pixels and attackable pixels.

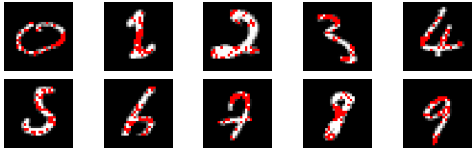

Fig. 6: Attackable pixels for 2-pixel attack highlighted in red. For each digit, the union of all pixels that are part of any 2-pixel attack is shown.

| digit | abs | co | coi |
|---|---|---|---|
| 0 | 7(0.9%) | 5(0.6%) | 1(0.1%) |
| 1 | 60(7.7%) | 23(2.9%) | 6(0.8%) |
| 2 | 119(15.2%) | 66(8.4%) | 19(2.4%) |
| 3 | 254(32.4%) | 169(21.6%) | 21(2.7%) |
| 4 | 3(0.4%) | 1(0.1%) | 1(0.1%) |
| 5 | 6(0.8%) | 1(0.1%) | 1(0.1%) |
| 6 | 2(0.3%) | 2(0.3%) | 2(0.3%) |
| 7 | 4(0.5%) | 2(0.3%) | 1(0.1%) |
| 8 | 142(18.1%) | 86(11.0%) | 19(2.4%) |
| 9 | 169(21.6%) | 98(12.5%) | 13(1.7%) |

TABLE V: Smallest number of important pixels to explore for the first 1-pixel attack for each image (digit) and each metric.

| digit | 0 | 1 | 2 | 3 | 4 | 5 | 6 | 7 | 8 | 9 |
|---|---|---|---|---|---|---|---|---|---|---|
| coi | 2 | 2 | 19 | 3 | 2 | 2 | 2 | 2 | 4 | 2 |

TABLE VI: Smallest number of important pixels to explore for the first 2-pixel attack for each image for the *coi* metric.

important pixels identified by the *coi* metric in DeepCheck$^{Imp}$ and select all $\binom{39}{2}$ = 741 unordered pairs that can be formed using the important pixels selected, and evaluate each pair to determine if it is attackable. Table III shows the number of attacks identified by DeepCheck$^{tPA}$ 2-pixel attack, and of those attacks the number that does not include any attackable pixel for 1-pixel attack. As expected, many 2-pixel attacks consist of a pixel that was 1-pixel attackable. However, several new attack pairs do not include any pixel that is attackable for 1-pixel attack are found. 4 out of 10 digits can be attacked to create multiple incorrect labels (alabel), e.g., digit 8 can be attacked using 3 different 2-pixel attacks to make the neural net incorrectly classify it as 1, 2, or 3.

For each digit, Figure 6 shows the union of all pixels in any 2-pixel attack to display their location. These pixels lie on or very close to the shape of the corresponding digit.

Overall, the use of symbolic execution enables finding both 1-pixel and 2-pixel attacks for each image.

### D. Important pixels and attackable pixels

Table IV presents a comparison of important pixels and attackable pixels for 1-pixel attack. For each image (digit), the column "#ap" shows the number of attackable pixels identified by DeepCheck$^{tPA}$ for 1-pixel attack. The group of 3 columns labeled 5% shows the percentage of attackable pixels that are in the top-5%, i.e., 39, of the important pixels identified by each of the 3 metrics of DeepCheck$^{Imp}$. The next 2 groups of 3 columns show the corresponding results for top-10% and top-30% of important pixels. At the top-5% and top-10% levels, *coi* outperforms the other two metrics in predicting attackability. For 3 of the 10 images, all attackable pixels lie in the top-5% of important pixels identified by *coi*. Moreover, for 6 of the 10 images, at least two-thirds of the attackable pixels are in top-5% for *coi*. Furthermore, for 8 out of 10 images, at least one-half of attackable pixels are in the top-5% for *coi*. At the top-30% level, the *abs* metric and *coi* metric perform similarly and have higher effectiveness than *co*, although the difference in the metrics reduces as % of important pixels increases. Overall, the results show that at the 5% and 10% levels, important pixels identified by *coi* likely contain majority of attackable pixels.

Table V shows for each image (digit) the smallest number of important pixels that must be explored before a 1-pixel attack is found for each of the three metrics. For all metrics and all images, no more than top one-third of the important pixels need to be checked to find an attack pixel. Moreover, for all metrics, less than 10 pixels need to be checked to discover an attack pixel for at least half of the images. The *coi* metric requires a maximum of 21 pixels to be checked across all the images, and for 4 images, the top most important pixel identified by *coi* is an attack pixel.

Table III already shows that forming pairs using just top-5% of the important pixels based on the *coi* metric allows several 2-pixel attacks on each image (digit). Table VI presents the smallest number of important pixels to explore to find the first 2-pixel attack for each image (digit) based on the *coi* metric. The worst case is for digit 2, where top-19 important pixels must be considered to find a 2-pixel attack. The best case happens for 7 out of 10 digits, where the top-2 important pixels allow DeepCheck$^{tPA}$ to create a 2-pixel attack; for these 7 cases, using DeepCheck$^{Imp}$ with *coi* metric presents an optimal strategy to find a 2-pixel attack. Moreover, for 9 out of 10 digits, a 2-pixel attack is found within just the top-4 important pixels, i.e., exploration of no more than 6 pairs.

Overall, important pixel identification using DeepCheck$^{Imp}$ plays an important role in focusing the exploration of DeepCheck$^{tPA}$ to find 1-pixel and 2-pixel attacks.

## VI. RELATED WORK

Recent independent work, developed concurrently with ours, proposes concolic testing for deep neural networks [27]. However their focus is on defining and achieving test coverage requirements, although their approach also produces adversarial images. In contrast we use symbolic execution for identifying important pixels and for specific 1-pixel and 2-pixel attacks, which target the same activation pattern as the original image; furthermore we use important pixels to focus the search for attackable pixels. Another difference is that we use off-the-shelf solvers. Other related recent techniques include formal methods [15] and testing [23], [30] for deep neural networks. However none of previous work uses formal methods for important pixel identification, or more generally for explainability in neural networks.

The rest of this section describes existing techniques related to attribution or explainability in neural networks and also existing techniques for adversarial example generation.

### A. Techniques for Attribution

Despite the wide-spread adoption of neural networks, most deep neural network classifiers are black-boxes. It is crucial to understand the reasons behind the predictions of these classifiers in order to build trust in the model. Therefore, a number of techniques have been explored in the area of generating explanations for predictions. *Attribution* is a specific class of approaches, mostly applicable to image classification applications, where the technique attempts to assign "relevance", "contribution" to each input feature or pixel towards the classification decision. We describe below broad categories of attribution approaches.

*Perturbation-based approaches* alter the value of every input feature individually by a specific amount [32], re-run the network on the input and then measure the difference in the output value. However, these techniques tend to be slow and the computation time increases with the number of features. *Gradient-based approaches* [25]) compute the attributions of every feature in a single forward and backward pass of the network on a given input. They compute the signed partial derivative of the output w.r.t each input variable and multiple it by the input value to determine the impact of that variable on the output. *Integrated-gradients* [28]) proposed an approach that take an average of the attributions calculate along a linear path from a baseline (user-defined) until the given input.

*Saliency maps* [26] consider the absolute value of the partial derivatives of the output w.r.t each input variable in order to identify pixels that can perturbed the least to observe a sizable change in the output value.

### B. Techniques for adversarial attack generation

It has been observed that state-of-the-art networks are highly vulnerable to *adversarial perturbations*: given a correctly-classified input $x$, it is possible to find a new input $x'$ that is very similar to $x$ but is assigned a different label [29]. Goodfellow et al. [14] introduced the Fast Gradient Sign Method for crafting adversarial perturbations using the derivative of the model's loss function with respect to the input feature vector. They show that NNs trained for the MNIST and CIFAR-10 classification tasks can be fooled with a high success rate. An extension of this approach applies the technique in an iterative manner [12]. Jacobian-based Saliency Map Attack (JSMA) [22] proposed a method for targeted misclassification by exploiting the forward derivative of a NN to find an adversarial perturbation that will force the model to misclassify into a specific target class. Carlini et. al. [5] recently proposed an approach that could not be resisted by state-of-the-art networks such as those using defensive distillation. Their optimization algorithm uses better loss functions and parameters (empirically determined) and uses three different distance metrics.

The DeepFool [21] technique simplifies the domain by considering the network to be completely linear. They compute adversarial inputs on the tangent plane (orthogonal projection) of a point on the classifier function. They then introduce non-linearity to the model, and repeat this process until a true adversarial example is found. Deep Learning Verification (DLV) [15] is an approach that defines a region of safety around a known input and applies SMT solving for checking robustness. They consider the input space to be discretized and alter the input using manipulations until it is at a minimal distance from the original, to generate possibly-adversarial inputs. DeepSafe [10] is an approach that first applies a label-guided clustering algorithm on inputs with known labels to identify input regions that can be expected to be consistently labeled. It then employs the Reluplex solver [16] to verify that the all possible inputs within a given region are assigned the same label by the network.

## VII. CONCLUSION

As deep neural networks become more and more commonly used in tasks of high importance, developing techniques that validate them becomes increasingly urgent. This paper introduced a new approach for validating neural networks based on the classic program analysis of symbolic execution. The key insight is to transform the network into an imperative program that is amenable to analysis using symbolic execution. Two analyses are presented: 1) to identify important pixels that can explain the classification decisions made by a neural network; and 2) to create 1-pixel and 2-pixel attacks by identifying pixels or pixel-pairs and computing their values so the neural network misclassifies the modified images. The two analyses apply in synergy and provide a more scalable approach to finding attacks. An experimental evaluation using the widely studied MNIST dataset demonstrates that the usefulness of symbolic execution in analyzing neural networks.

## APPENDIX

Table VII highlights in green the top-30% (i.e., 235) important pixels identified by DeepCheck$^{Imp}$ for each of the three metrics *abs*, *co*, and *coi* for each image (digit).

| digit | abs | co | coi |
|---|---|---|---|
| 0 | | | |
| 1 | | | |
| 2 | | | |
| 3 | | | |
| 4 | | | |
| 5 | | | |
| 6 | | | |
| 7 | | | |
| 8 | | | |
| 9 | | | |

TABLE VII: Top-30% (i.e., 235) of important pixels (green) identified by DeepCheck$^{Imp}$ for *abs*, *co*, and *coi*.